\documentclass[conference]{IEEEtran}
\usepackage[noadjust]{cite}  
\usepackage{graphicx}
\usepackage{wrapfig}
\usepackage{lipsum}
\usepackage{dblfloatfix}
\usepackage{fixltx2e}
\usepackage{color}
\usepackage{nameref}
\usepackage{amsmath}
\usepackage{graphicx}
\usepackage{varioref}
\usepackage{booktabs}  % nice looking tables
\usepackage{siunitx}
\usepackage{numprint}
\usepackage{filecontents}
\usepackage[bookmarks=false]{hyperref}
\ifCLASSINFOpdf
\else
\fi
\IEEEoverridecommandlockouts 
\begin{document}
\title{EchoWear: Smartwatch Technology for Voice and Speech Treatments of Patients with Parkinson's Disease\\ }
\author{\IEEEauthorblockN{Harishchandra Dubey\thanks{{\color{blue}This material is presented to ensure timely dissemination of scholarly and technical work. Copyright and all rights therein are retained by the authors or by the respective copyright holders. The original citation of this paper is:
H. Dubey, J. C. Goldberg, M. Abtahi, L. Mahler, K. Mankodiya, EchoWear: Smartwatch Technology for Voice and Speech Treatments of Patients with Parkinson's Disease, choWear: smartwatch technology for voice and speech treatments of patients with Parkinson's disease. In Proceedings of the conference on Wireless Health (WH 2015), National Institute of Health, Bethesda, Maryland, USA, ACM, Article 15 , 8 pages. DOI: http://dx.doi.org/10.1145/2811780.2811957.  }}\IEEEauthorrefmark{1}, Jon C. Goldberg\IEEEauthorrefmark{1}, Mohammadreza Abtahi\IEEEauthorrefmark{1}, Leslie Mahler\IEEEauthorrefmark{2}, Kunal Mankodiya\IEEEauthorrefmark{1}}
\IEEEauthorblockA{\IEEEauthorrefmark{1}Department of Electrical, Computer and Biomedical Engineering, University of Rhode Island, Kingston, RI 02881, USA\\ \IEEEauthorrefmark{2}Department of Communicative Disorders, University of Rhode Island, Kingston, RI 02881, USA\\
dubey@ele.uri.edu, lmahler@uri.edu, kunalm@uri.edu}}
%story for this paper:
\maketitle
\begin{abstract}
About 90 percent of people with Parkinson’s disease (PD)
experience decreased functional communication due to the
presence of voice and speech disorders associated with dysarthria
that can be characterized by monotony of pitch (or fundamental
frequency), reduced loudness, irregular rate of speech, imprecise
consonants, and changes in voice quality. Speech-language
pathologists (SLPs) work with patients with PD to improve
speech intelligibility using various intensive in-clinic speech
treatments. SLPs also prescribe home exercises to enhance
generalization of speech strategies outside of the treatment room.
Even though speech therapies are found to be highly effective in
improving vocal loudness and speech quality, patients with PD
find it difficult to follow the prescribed exercise regimes outside
the clinic and to continue exercises once the treatment is
completed. SLPs need techniques to monitor compliance and
accuracy of their patients’ exercises at home and in ecologically
valid communication situations. We have designed EchoWear, a
smartwatch-based system, to remotely monitor speech and voice
exercises as prescribed by SLPs. We conducted a study of 6
individuals; three with PD and three healthy controls. To assess
the performance of EchoWear technology compared with highquality
audio equipment obtained in a speech laboratory. Our
preliminary analysis shows promising outcomes for using
EchoWear in speech therapies for people with PD.

\textbf{Keywords- Dysarthria; knowledge-based speech processing; Parkinson’s
	disease; smartwatch; speech therapy; wearable system.}
\end{abstract}
\IEEEpeerreviewmaketitle
%%\vspace{-1mm}
\section{Introduction}
\begin{figure}[!t]
	\centering
	\includegraphics[width=240bp]{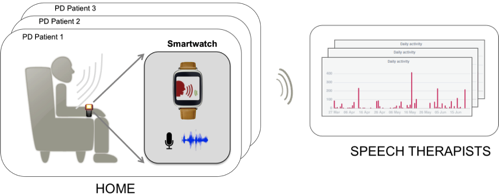}
	\caption{A concept of the EchoWear system.}
	\label{fig1}
\end{figure}
Parkinson disease (PD) is the second most common
neurodegenerative disorder of mid-to-late life in developing and
developed countries~\cite{e1_de2006epidemiology}. Approximately 4 million people
worldwide were diagnosed with PD in 2005 and that number
projected to go beyond 9 million by 2030~\cite{e2_dorsey2007projected}. The characteristic
motor disorder that defines PD includes rigidity, slowness of
movement (bradykinesia), and hypokinesia. Speech problems are
common in people with PD and it has been estimated that 70-90%
of patients reported speech impairments after the onset of PD~\cite{e3_hartelius1994speech,e4_ho1999speech}. Patients with PD experience a combination of speech impairments including; reduced vocal loudness~\cite{e5_ho2001motor}; a breathy or
harsh voice quality~\cite{e6_baumgartner2001voice}; imprecise consonants and distorted vowels~\cite{e7_sapir2007effects}; and reduced voice pitch (fundamental frequency) variation~\cite{e8_forrest1989kinematic} collectively called hypokinetic dysarthria~\cite{e9_duffy2013motor}.

Speech treatments are effective to enhance speech intelligibility,
voice quality and confidence of patients with PD to communicate~\cite{e7_sapir2007effects}. However, it is challenging for patients to maintain long-term
benefits of treatment since PD progresses uniquely in each patient.
Therefore, SLPs design a personalized approach for each patient
to set individual speech goals in treatment. It is difficult for SLPs
to accurately assess whether patients adhere to the prescribed
therapy at home and in functional communication situations
outside of the clinic. Since PD may also affect cognitive abilities
including memory, patients may not remember precise details of
the therapy exercises and the recommended exercise schedule.
Hence, SLPs seek an efficient and effective solution to remotely
monitor the speech of their patients.

We have developed a smartwatch-based system, "EchoWear"
(shown in Figure~\ref{fig1}), to collect data on various attributes of speech
exercises performed by patients with PD outside of the clinic. In
this paper, we provide results of research conducted with patients
with PD and healthy adults to validate the performance of
EchoWear to record quality speech data. In the subsequent
sections, we will describe the architecture of EchoWear that
enables recording, processing and communication of wearers' speech data. In-depth comparisons between the data from EchoWear and audio equipment used by SLPs are discussed to
demonstrate the reliability and validity of modern smartwatch
technology for its use as a tele-recording device.
%%
%\vspace{-2mm}
%\vspace{0mm}
%
\section{Background \& Related Works}
\subsection{Speech Disorders in People with PD}
The symptoms of PD are associated with alterations in basal
ganglia circuitry due to decreased in dopamine in the substantia
nigra pars compacta~\cite{e10_damier1999substantia,e11_ahlskog2007beating,e12_braak2004stages}. However, the neural mechanisms underlying the effects of dopamine loss and its impact on speech
and voice are not well understood. Physiological abnormalities
associated with speech and voice changes in people with PD
include reduced vocal fold adduction and asymmetrical patterns of
vocal fold vibration~\cite{e13_perez1996parkinson,e14_smith1995intensive}; reduced neural drive to laryngeal muscles~\cite{e15_baker1998thyroarytenoid}; poor reciprocal suppression of laryngeal and
respiratory muscles~\cite{e16_vincken1984involvement}; and a reduction in respiratory muscle
activation patterns~\cite{e17_solomon1993speech} all of which contribute to the perceptual
feature of significantly decreased loudness in people with PD.
Motor speech characteristics of rigidity, weakness, bradykinesia
and hypokinesia do not completely account for the speech
abnormalities associated with PD. Additional non-dopaminergic
mechanisms such as sensory deficits in the internal monitoring of
amplitude and maintaining amplitude of speech movements and
volume of speech are significant factors that also contribute to
decreased loudness, imprecise articulation, and limited pitch
variation~\cite{e20_sapir2011intensive,e21_desmurget2004basal,e22_schneider1986deficits}.
\subsection{Speech Therapies in PD}
Speech therapy is an important element of treatment for patients
with PD. Traditional speech therapy typically involves multiple
speech system targets such as voice, rate, articulation, and
respiration~\cite{e23_ramig2001changes}. For example, the Lee Silverman Voice Treatment
(LSVT LOUD) has been used as an effective therapy in the short
term and long term to improve speech loudness and quality in
people with PD by targeting voice~\cite{e23_ramig2001changes,e24_ramig2001intensive}. LSVT LOUD is
intensive (4 days a week or 16 sessions in one month) and
systematic in training the vocal loudness~\cite{e25_fox2012lsvt}. Regardless of the
specific treatment approach, patients have to participate in
treatment proactively by performing home exercises prescribed by
their SLPs~\cite{e26_cherney2012telerehabilitation}. Regular home exercises and using speech
strategies in functional communication situations are as important
as the intensive in-clinic training given by SLPs. Acoustic
analysis of diadochokinesis for dysarthric speech was proposed
and validated in~\cite{e39_tjaden2002characteristics}. The temporal features proved to be better
than energy features for discriminating dysarthria secondary to
multiple sclerosis, dysarthria secondary to PD, and healthy
controls. The authors performed acoustic analyses based on
duration and Bark-scaled F1-F2 values of the vowels. The PD
participants did not show an effect of density on dispersion for
high-frequency words~\cite{e40_watson2008parkinson}. Induced variability in F2 trajectories
for different speaking rates in patients with PD and healthy
controls is discussed in~\cite{e41_tjaden1998speaking}. It was demonstrated that speaking
rate did not have a consistent influence on F2 onset frequency for
both healthy controls and patients with PD.
\subsection{Technologies for PD Speech Treatments}
Recently, increasing numbers of SLPs have adopted telehealth or
tele-rehabilitation services involving information and
communication to enhance treatment methods~\cite{e26_cherney2012telerehabilitation}. Online speech therapy or tele-practice leverages internet-connected computers
with a webcam, speakers and a microphone to form a clinical
arrangement where the patient and an SLP can communicate faceto-
face over the Internet from different locations~\cite{e27_theodoros2008telerehabilitation}. For
example, the LSVT LOUD companion software allows SLPs to
access their patients’ homework and exercises completed outside
the clinic environment~\cite{e28_goetz2009testing}.

EchoWear leverages modern smartwatch technology that comes
with a variety of sensors, an interactive touch screen, and an
ability to exchange the data and information with smartphones for
the purpose of monitoring speech exercises at home and in
functional communication situations. The smartwatch is used as a
wearable sensor worn on the wrist of patients to tele-monitor how
they follow up with speech therapy at home. The use of
smartwatches for such tele-monitoring applications demands the
design of a reliable architecture such as architecture of EchoWear
to first validate its performance through controlled in-clinic
validation trials.
\begin{figure*}[!t]
	\centering
	\includegraphics[width=480bp]{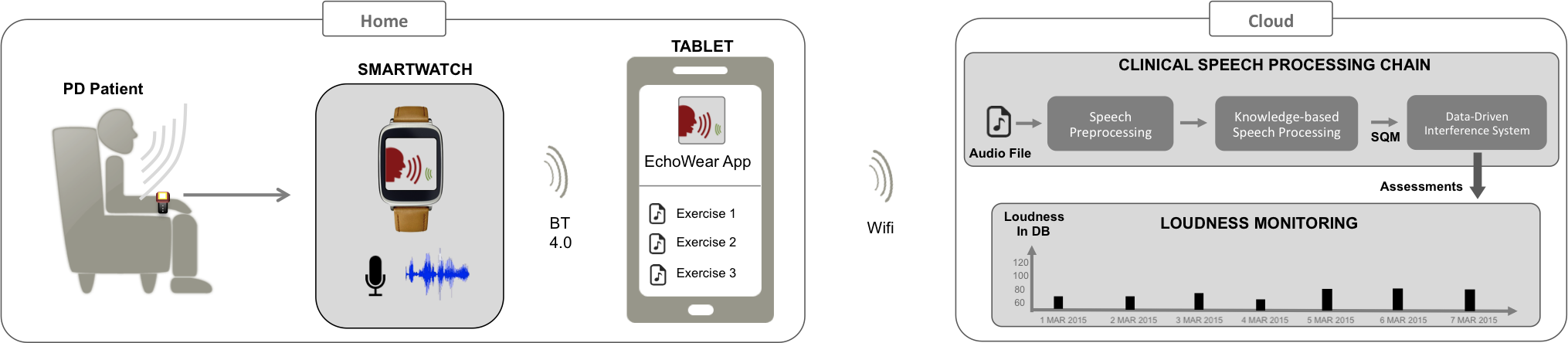}
	\caption{System architecture of EchoWear.}
	\label{fig2}%
\end{figure*}
\section{EchoWear – A Wearable System for Speech Treatments}
EchoWear is a wearable speech monitoring system to leverage
sensing and communication capabilities of modern smartwatches
to generate a dynamic structure of monitoring speech exercises in
patients with PD. The system architecture of EchoWear is divided
into three elements as described below.
\subsection{Smartwatch System}
\begin{figure}[!t]
\centering
\includegraphics[width=240bp]{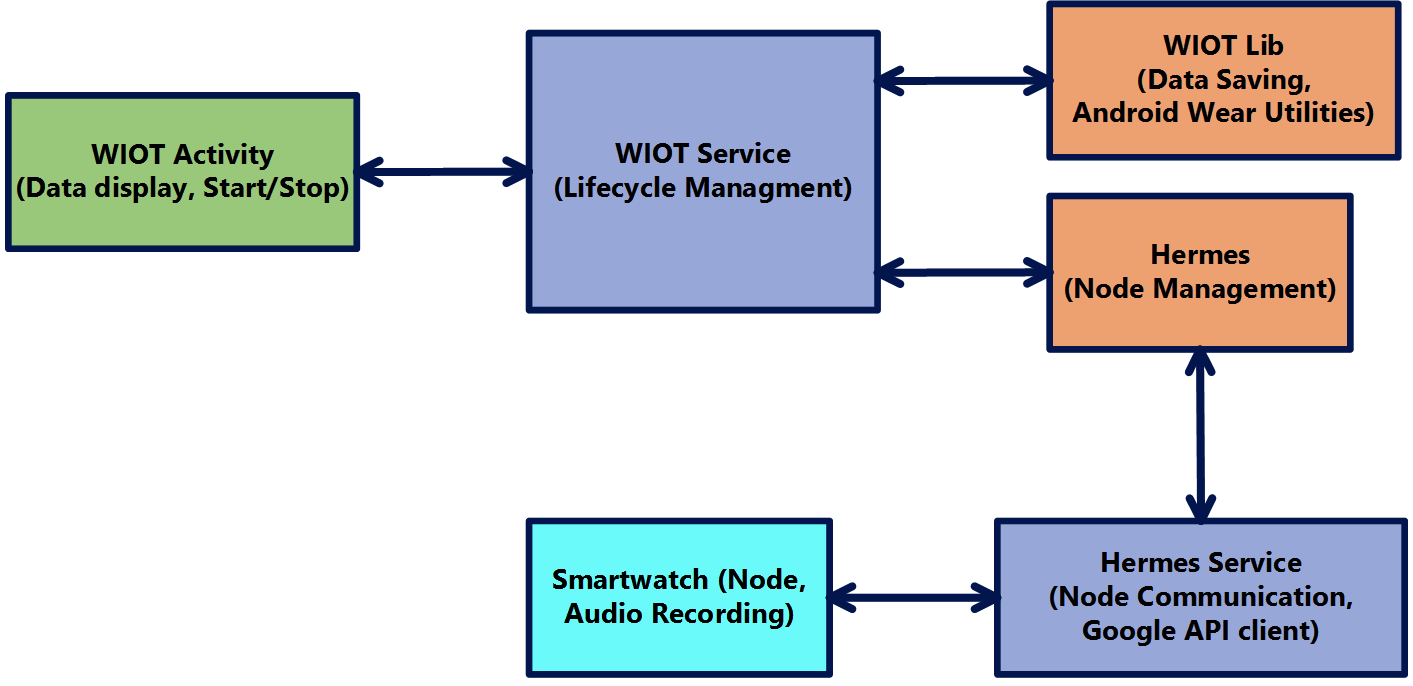}
\caption{Interplay between the smartwatch and the tablet.}
\label{fig3}%
\end{figure}

The reason we call it a smartwatch system is that the smartwatch
is not a standalone device. It works in conjunction with a
smartphone or a tablet for short-range communication to provide
interplays such as extended notifications of messages and phone
calls, voice command control, and physical activity monitoring
including step counting. Essentially, the smartwatch is considered
an extended part of the smartphone system and provides
opportunity for users to respond instantaneously to activities on
their smartphones. As shown in Figure~\ref{fig2}, EchoWear uses the
combination of a smartwatch and a smartphone for speech
therapy. PD patients wear modern smartwatches running Android
Wear OS. The smartwatch receives a control from a nearby
smartphone (tablet) signaling the recording process. The speech
data is received through the smartwatch’s built-in microphone
followed by filtering to remove background noise using the
Android API for audio. The recording frequency is set to 44.1
kHz with 16-bit precision. We have developed a Wearable
Internet of Things (WIOT) framework that allows Android
devices to connect seamlessly to nearby-placed wearable devices
such as smartwatches~\cite{e38_hiremath2014wearable}. Once the Android tablet initiates the
recording process, the smartwatch continuously streams the data
through the use of the Bluetooth 4.0 protocol, in conjunction with
Google’s Wearable Message API. The data is sent from the
smartwatch in a 2 kB package, until the recording process has
been completed. During the receiving, the data is buffered into an
output stream on the smartphones internal storage. Once the
process has completed, the RAW audio data is re-read, the WAV
header is added, and is then saved in a format compatible with most audio players.
%\vspace{-2mm}
%
The WIOT framework designed for EchoWear correctly aligns
itself with the standard Android application lifecycle. To maintain
compatibility and stability, the components of WIOT are split up
into key components. As shown in Figure~\ref{fig3}, the framework is
split into an activity, a service, WiotLib, Hermes, and finally the
smartwatch. Starting from what the user sees, the activity is the
forefront of the framework. This activity displays the data
collection process, and controls for maintaining the data. The
activity directly speaks to the service. The service is responsible
for the lifecycle management of both the activity and the
collection of APIs used for the research. WiotLib acts as a
collection of static methods to make the programming part easier,
allowing us to reuse standard code across multiple projects.
Hermes, an in-house messaging service, was designed to handle
communication between the service and the smartwatch. Hermes
allows the addition of multiple smartwatches, and receives a
variety of different types of data. Hermes also lets us maintain the
smartwatch lifecycle, and keep track of battery life, and prevent
the watch from entering commercial modes, such as battery
saving features. Finally, we are running a service on the
smartwatch that responds to Hermes. This service is used to
communicate with the onboard hardware, specifically the
microphone. The tablet sends the speech data obtained from the
smartwatch to the cloud server. We have a speech analysis engine
in the cloud that process the speech signal as discussed in the next
section.
%
%\vspace{-1mm}
%%\vspace{-1mm}
%
\subsection{The Cloud for Speech Analysis}
The cloud stores the speech data obtained from the tablet during
speech exercises of PD participants and processes the speech data.
It has two main units, namely an analysis unit and a visualization
unit. The analysis unit computes the speech quality metric
(SQMs) and the visualization unit displays the results on user
interface. The audio files accumulated in the cloud are analyzed
by a knowledge-based clinical speech processing chain (CLIP) to
get clinically relevant metrics like loudness and frequency. CLIP
is a modular software chain with the possibility to incorporate
other clinically relevant metrics such as jitter, shimmer, sensory
pleasantness, and dysphonia measures.
\subsubsection{Clinical Speech Processing Chain (CLIP)}
\label{sec:s3p2p1}
CLIP is a flexible software system in the cloud that computes
perceptual speech quality metrics (SQMs). CLIP consists of
several sub-systems as shown in Figure~\ref{fig2}. The speech signal is
pre-processed to make it suitable for acoustic analysis. The
knowledge-based speech processing block takes the processed
speech and computes SQMs based on mathematical models of
human auditory perception. The speech signals can have two
types of sounds, i.e., voiced sounds (that can be vowels or nasal
sounds) and unvoiced sounds (fricatives and plosive sounds that
are consonants). The SQMs computed in this paper are listed in Table~\ref{table1}. The SLPs use SQMs to monitor the speech quality of
participants and to infer if the participant has improved by
performing vocal exercises at home. The final block in CLIP is
the data-driven inference system that uses large amounts of SQMs
computed over the time to provide automatic health reports to
SLPs and/or participants. This block is essentially a machine
learning system that adapts itself for each participant and SLP to
provide personalized speech treatment for PD. The ultimate goal
of CLIP within EchoWear is to provide a fully automated,
intelligent and flexible enhanced speech treatment for PD
participants.
\section{Method}
\subsection{Participants}
Seven participants were recruited for this study. Four participants
diagnosed with PD were recruited from the Department of
Communicative Disorders in the University of Rhode Island. One
of the PD participants withdrew from the study because of illness.
Three out of six participants were diagnosed with PD and the time
since diagnosis was from 3 years up to 25 years. Three
participants without PD served as healthy controls. The
participants $S_{1}$, $S_{2}$, and $S_{3}$, were diagnosed with PD and participants
$S_{4}$, $S_{5}$ and $S_{6}$ were healthy controls.
\subsection{Protocol}
The participants were asked to perform three speech tasks. Task 1
(t1) was a vowel prolongation task, in which participants were
asked to sustain the vowel “ah” for as long as possible for a total
of three repetitions. Task 2 (t2) and Task 3 (t3) were developed to
record high and low pitches. Participants were asked to start
saying "ah" at their talking pitch and then go up or down in pitch
and hold it for 5 seconds and repeat it for three repetitions. All the
instructions related to each task were shown on a screen in front
of the participants and were explained before starting each task.
\subsection{Experimental Setup}
\label{sec:s4p3}
\begin{figure}[!t]
	\centering
	\includegraphics[width=240bp]{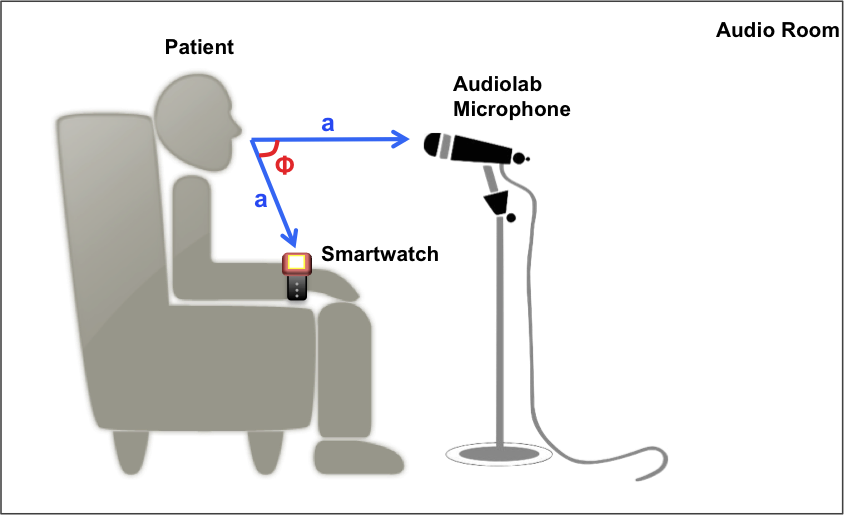}
	\caption{Acoustic scenario for speech data collection.}
	\label{fig4}%
\end{figure}

Evaluations took place in an IAC sound-treated booth at the
University of Rhode Island Speech and Hearing Center. The
recording environment is shown in Figure~\ref{fig4}. Each participant was
seated in a chair and wore an Android smartwatch––Asus
Zenwatch while simultaneously collecting data using audio
recording technology~\cite{e33_device}. A head-mounted microphone (model
Isomax B3) was placed at a distance of 8 cm from the mouth and
even with the participant’s mouth. A sound level meter (SLM;
Bruel \& Kjaer Type 2239) was placed at a distance of 40 cm from
the participant’s mouth. The head-mounted microphone and SLM
signal were digitized and directly sent to the computer (Toshiba
Qosmio). Speech was sampled at 44 kHz using Goldwave
software. Evaluations were also recorded using a Cannon FS400
camcorder.
The participants were asked to maintain the microphone as well as
the smartwatch at the same distance from the throughout the
recording. The rectangular enclosure represents the clinical room
where experimental data were collected. The speech signal from
the mouth can follow several reflected paths in addition to the
direct path to reach the smartwatch or the microphone. However,
since the microphone and the smartwatch were at different
orientations (different positions), the reflected path for each case
was different. Accounting for the environmental variables such as
reverberation and room impulse response was out of the scope in
this validation study. Since the recording took place in a sound
treated booth, calibration based on room impulse response was not
needed. The aim of this experiment was proof-of-concept for the
smartwatch compared with traditional speech recording methods
in a controlled acoustic environment. The speech amplitude has
shallow dependence on orientation and distance from mouth,
hence for small movements made by participants, the deviations
were insignificant as shown in Section~\ref{sec:s5}.
%
%\vspace{0mm}
\begin{figure}[!t]
	\centering
	\includegraphics[width=240bp]{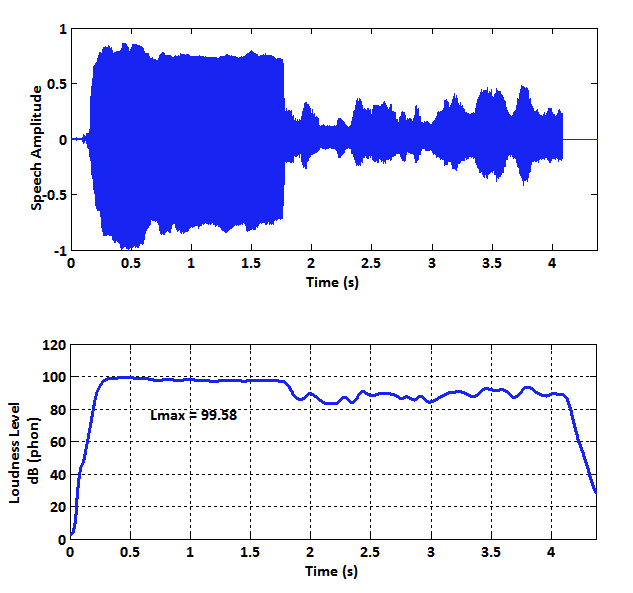}
	\caption{Speech signal and corresponding instantaneous
		loudness level in dB (Phon) for participant $S_{1}-t_{1}$ (baseline).
		The loudness level has a strong dependence of amplitude as
		depicted here. It has shallow dependence on frequency content
		and time duration of speech signal.}
	\label{fig5}%
\end{figure}
\subsection{Proof-of-Concept Trial}
We received an approval (ref no: 682871-2) from the Institutional
Review Board to conduct our experiments involving individuals
with PD and healthy controls. Participants read and signed the
consent form at the time of data collection. All the participants
were introduced to the new technology involved in this trial. The
trials were conducted by a certified SLP at URI. Participants had
the option to terminate the ongoing trial at any time.
%
%\vspace{0mm}
\begin{figure*}[!t]
	\centering
	\includegraphics[width=480bp]{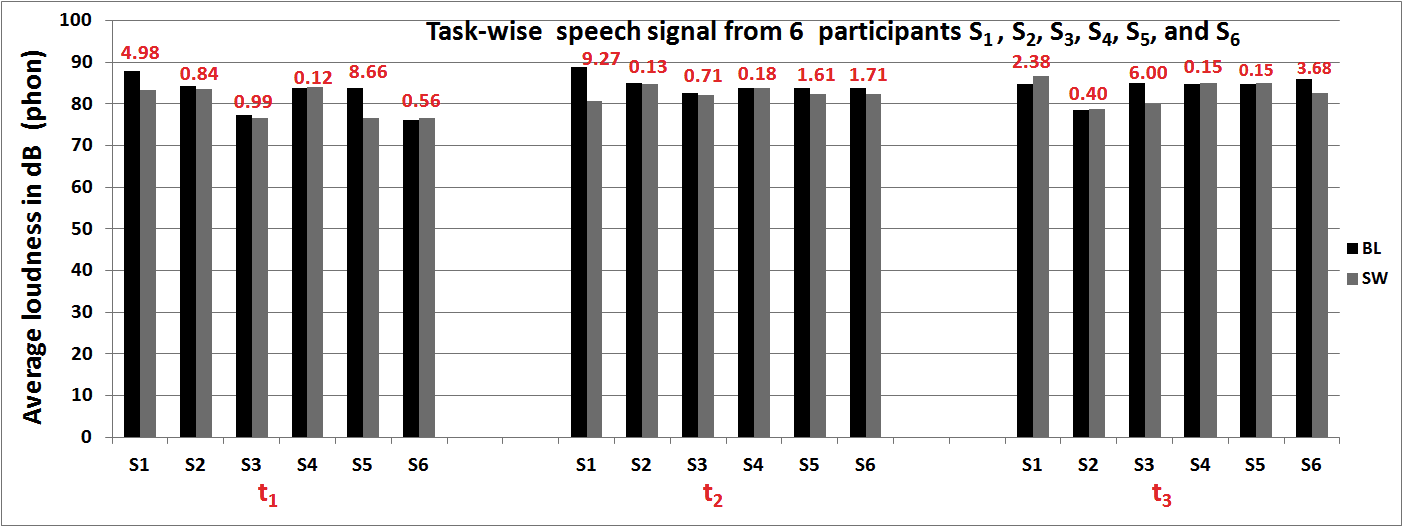}
	\caption{Comparison of average loudness level in dB (Phon) for baseline (BL) and smartwatch (SW) speech signals (Numbers above the bars represent percent deviation from the smartwatch data compared to the baseline).}
	\label{fig6}%
\end{figure*}
\begin{figure}[!t]
	\centering
	\includegraphics[width=240bp]{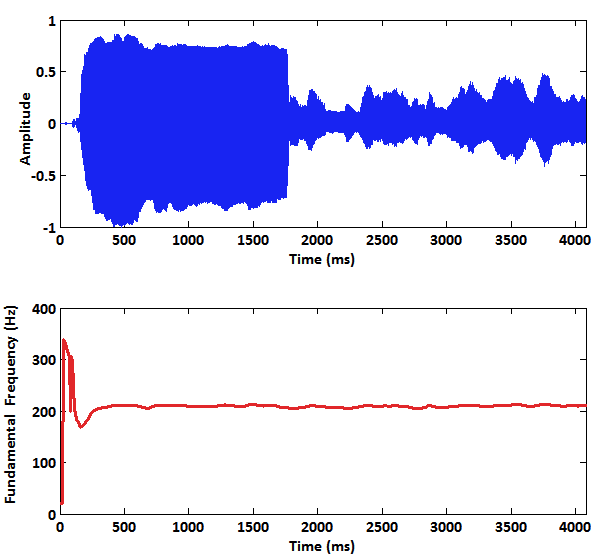}
	\caption{Fundamental frequency contour for speech signal
		$S_{1}-t_{1}$ (baseline). The first few samples of speech signal has very
		low amplitude (unvoiced speech) that corresponds to very
		high instantaneous frequency (Hz). The unvoiced speech does
		not cause perception of pitch. The instantaneous frequency
		corresponding to unvoiced speech is not accounted for
		computation of average fundamental frequency.}
	\label{fig7}%
\end{figure}
\begin{figure*}[!t]
	\centering
	\includegraphics[width=480bp]{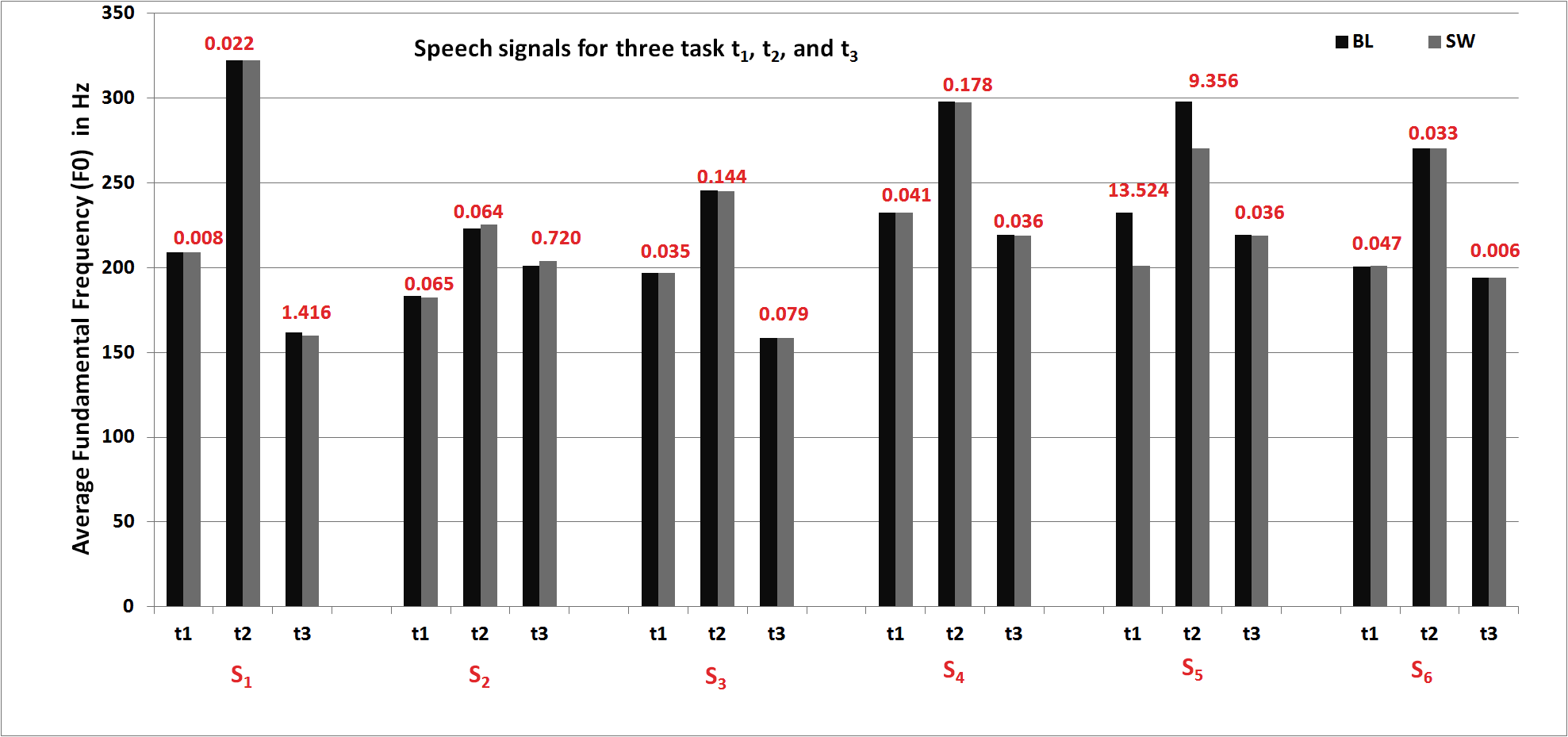}
	\caption{Comparison of average fundamental frequency (F0) in Hz for baseline (BL) and smartwatch (SW) speech signals (Numbers above the bars represent percent deviation from the smartwatch data compared to the baseline).}
	\label{fig8}%
\end{figure*}
\begin{table*}[t]
	\centering
	\caption{List of speech quality metrics (SQMs)}
	\begin{tabular}{*{2}{|c|}}
		\hline
		SQM & Definition \\ 	\hline
		Average loudness level in dB (Phon) & Average of the instantaneous loudness
		level in dB (Phon) \\ 	\hline
		
		Average fundamental frequency (Hz) & Average of the fundamental frequency
		(F0) contour \\ 	\hline
	\end{tabular}
	\label{table1}
\end{table*}
\section{Results \& Discussions}
\label{sec:s5}
The loudness and fundamental frequency ($F_{0}$) were two primary
SQMs for assessment of speech. We will discuss the mathematical
foundations of speech processing needed for computing these
SQMs and validate the accuracy of the smartwatch technology in
terms of these SQMs with respect to baseline microphone data.
We will also discuss the practical limitations that cause variations
in SQMs computed using the smartwatch instead of baseline
microphone.
\subsection{Speech Pre-processing}
\label{sec:s5p1}
The speech signals were recorded using both a smartwatch and a
microphone. The acquired speech signal was contaminated with
background noise, and SLP’s voice so we edited the speech signal
to remove the SLP’s voice (interruptions). After editing, we use
the spectral subtraction for reducing the background noise~\cite{e30_berouti1979enhancement}.
The spectral subtraction is a simple method for noise reduction
based on the assumption of stationary white Gaussian noise
uncorrelated with the clean speech signal. The short-time noise
spectrum is computed during the ‘silence’ frames and later
averaged and smoothed in frequency domain. The magnitude of
the smoothed estimates of the short-time time noise spectrum is
subtracted from the magnitude of short-time spectrum of noisy
speech signal to give the magnitude of the enhanced speech
spectrum. The phase of the noisy speech signal is used with the
magnitude of the enhanced speech spectrum to synthesize the
discrete-time enhanced speech signal by inverse Fourier
transform. For results reported in this paper, we used the method
described in~\cite{e31_martin2001noise} and~\cite{e32_gerkmann2012unbiased} to obtain accurate estimates of noise
spectrum.

The speech signal is short-time stationary with a period of 25 to
40 msec. Consequently, a common practice in speech processing
is to divide the speech signal into short time-frames of order
25msec. The overlapping time-frames were multiplied with a
Hanning window to prevent the spectral leakage. The windowed
time frames are processed by fast Fourier transform (FFT) to give
the short-time spectrum of speech signal~\cite{e34_cooley1969fast}. For monitoring the
PD participants, SQMs that quantify the perception of the speech
signal by the human auditory system are needed. These SQMs are
derived from short-time speech spectrum with knowledge of
auditory models. The SLPs use average loudness and average
fundamental frequency ($F_{0}$) that are derived from short-time
spectrum of speech signal.
\subsection{Loudness}
\textsl{Loudness} is the perceptual correlate of intensity of the speech
signal. The loudness computation was based on various auditory
models suitable for different types of sounds. We used the
Zwicker‘s method for loudness computation valid for time
varying sound for PD speech~\cite{e35_rennies2010comparison,e36_zwicker2013psychoacoustics}. This method is standardized
as DIN 45631/A1 (2008). The human auditory perception is
frequency selective. Its frequency selectivity is captured by a nonlinear
scale known as the Bark-scale. The critical-band rates
(defined by the bark scale) play an important role in loudness
computation. The specific loudness of a frequency-bin (particular
frequency) is denoted as N!, and measured in Sone/Bark.
Loudness, $N$ (in unit Sone) is the integral of $N_{0}$, over all criticalband
rates. Mathematically, it is written as
\begin{equation}
N = \int_{n=0}^{24 Bark} N_{0} \cdot dz
\end{equation}

Typically, the step-size dz is 0.1 and sum is taken over all criticalband
rates. Sone and Phon are two different units of loudness~\cite{e36_zwicker2013psychoacoustics}. In this paper, we use Phon (in dB) as unit of loudness level.
We denote the six participants as $S_{1}$, $S_{2}$, $S_{3}$, $S_{4}$, $S_{5}$, and $S_{6}$ and
there are three tasks denoted by $t_{1}$, $t_{2}$ and $t_{3}$, respectively. Figure~\ref{fig5} shows a time domain speech signal and corresponding instantaneous loudness level in dB (Phon). The loudness depends on amplitude, frequency and duration of the speech segment. The
strong dependence of loudness on amplitude of speech signal is
clearly visible from this figure. The instantaneous loudness level
is high till 1.75 seconds where the amplitude of speech signal is
comparatively higher and after that it decreases. The
instantaneous loudness level follows the amplitude of the speech
signal. Loudness  has shallow dependence on frequency and
time duration of speech signal~\cite{e36_zwicker2013psychoacoustics}.
We analyzed two types of speech data: one from the baseline
microphone (BL) and another from the smartwatch (SW). We
computed the average loudness level (in dB) for all speech signals
as shown in Figure~\ref{fig6}. As depicted in this figure, the difference
between the two measurements is less than 5 percent except $S_{5}-t_{1}$,
$S_{1}-t_{2}$, and $S_{3}-t_{3}$ where it is 8.66, 9.27 and 6.00 percent
respectively. These percentage values are the percent deviation of
SW measurements taking BL measurements as reference. Both of
the speech signals are processed by the same method as discussed above. The variation in the two loudness values was due to the
fact that slightly different versions of the original speech signal
are acquired by the smartwatch and the microphone as discussed
in Section~\ref{sec:s4p3} and Section~\ref{sec:s5p1}. Thus, the smartwatch data can be
used to compute a reliable estimate of speech loudness. Due to
orientation (angular) differences between the SW and BL, the
amplitudes of SW and BL speech signals are slightly different.
The BL signals are dual channel and both channels were averaged
to form a mono channel speech signal before the pre-processing
step. The SW signals are mono channel.
\subsection{Fundamental Frequency}
%\vspace{0mm}
%
Voiced sounds are periodic, and possess information about pitch.
However unvoiced sounds are random white noise and do not
possess information about pitch. Voiced sound is produced by the
rapid vibration of the vocal folds. Pitch is the perceived frequency
of a sound and is approximately given by the fundamental
frequency ($F_{0}$). Typically, pitch varies from 80 to 160 Hz for
male and from 160 to 400 Hz for female. The pitch can be
approximately computed from peaks in the log-magnitude
spectrum of the speech. The pitch depends on the languages as
well as the person. The audible frequencies for humans lie
between 20 Hz and 20 kHz. However, most of the speech power is
typically contained in a range of 1.5 to 3.4 kHz~\cite{e29_ogunfunmi2010principles}. 

The speech is sampled at 44.1 kHz and stored with 16-bit precision within
EchoWear resulting in high fidelity speech signal. Physically, $F_{0}$
is related to the rate of vibration of vocal folds. The variation in
$F_{0}$ reflects the changes in intonation of speech, i.e., rise and fall of
speech while speaking. $F_{0}$ is the second most important SQM after
loudness. The inverse of the time-period of the speech signal is
the fundamental frequency. The energy in the speech spectrum is
concentrated mostly at integer multiples of fundamental frequency
($F_{0}$). The sinusoidal components of the speech signal with
frequencies above the fundamental frequency are called the
harmonics. For $F_{0}$ estimation [50 Hz, 500 Hz] is chosen as the
search range that is higher than the expected physical values for
voice as discussed in Section~\ref{sec:s3p2p1}. For each short time-frame of
speech signal we compute the instantaneous $F_{0}$ . The time varying
$F_{0}$  for speech signal is known as $F_{0}$  contour. The difference
between the highest and lowest frequency in $F_{0}$  contour is a
measure of pitch range. There are several algorithms for $F_{0}$ 
estimation. We used SWIPE because it is a frequency domain
algorithm for pitch detection and gives the best results for our
speech data. SWIPE is a sawtooth waveform inspired pitch
estimator for speech and music developed in~\cite{e37_camacho2008sawtooth}. It tends to find
the frequency that maximizes the average peak to valley distance
at harmonics of that frequency. It avoids taking the logarithm of
the spectrum and applies monotonically decaying weight to the
harmonic components. The logarithm operation leads to numerical instabilities for spectral nulls that are avoided in SWIPE. The frequency is transformed into the equivalent rectangular
bandwidth (ERB) scale before multiplying the weights to improve
the performance of SWIPE algorithm. The ERB scale mimics the
frequency sensitivity of the cochlea (in the inner ear) of human
auditory system. ERB leads to more accurate $F_{0}$ estimates. The
average $F_{0}$ is computed from $F_{0}$ contour for both BL as well as
SW data. Figure~\ref{fig7} shows a speech signal and corresponding
estimates of instantaneous $F_{0}$ (in Hz) by the SWIPE algorithm.
We can see that first few samples of speech have very low
amplitude (unvoiced sound) that does not have pitch as discussed
in Section~\ref{sec:s3p2p1}. Consequently, the pitch estimates corresponding
to unvoiced speech (low amplitude) are not relevant and hence not
accounted for computing average $F_{0}$. Figure~\ref{fig8} shows the average
$F_{0}$ (averaged over $F_{0}$ contour corresponding to voiced sound, i.e.,
excluding the first few samples that are very high) for both BL
and SW data. We can see that both BL as well as SW data give
almost the same average $F_{0}$. The highest deviation from baseline
is 9.35 percent for $S_{5}-t_{2}$. Small variations occur in average F0 due
to slightly different speech signals from baseline microphone and
smartwatch as discussed in the previous section. These variations
are insignificant with respect to monitoring of PD progression by
SLPs as we are interested in comparative values of SQMs over
different days of speech exercise. Hence, for estimation of
average $F_{0}$ SWIPE algorithm can be used with smartwatch in
EchoWear framework.
% %
%\vspace{-2mm}
\section{Conclusions}
The majority of individuals with PD face dysfunctional speech
and seek clinical help from SLPs to improve their speech
intelligibility and voice quality. Patients often participate in
intensive speech therapies to improve communication
effectiveness. A major challenge to treatment is carryover and
generalization of speech strategies outside the clinical
environment. Therefore, SLPs seek new ways assess exercise
compliance and monitor speech in environmentally relevant
communication situations. The current study presented the design
and validation of EchoWear, a smartwatch-based system for
speech treatments and demonstrated that SLPs could use the
smartwatch data and process it to obtain valid and reliable data for
tele-monitoring the speech of patients with dysarthria. We
recruited 6 individuals with and without PD to validate the
reliability of EchoWear. In this work, we analyzed loudness and
fundamental frequency as measures of speech characteristics. The
results suggest that EchoWear data were comparable to data
collected using traditional speech recording methods, even though
EchoWear used a mono channel audio signal unlike the dual
channel microphone system used by SLPs. The data support
EchoWear as a reliable framework to collect speech data from inhome
speech exercises. It has the potential to provide SLPs with a
new tool for monitoring speech during exercises and functional
communication to maximize generalization of speech goals
outside the clinical setting. Further research is required for us to
customize EchoWear for both patients and SLPs such that patients
can be trained to use smartwatches daily and SLPs can follow up
with their patients with the data analytics. In the future, the
EchoWear can be extended to other populations of people with
dysarthria such as those diagnosed with stroke, cerebral palsy,
traumatic brain injury or Down syndrome.
\section*{Acknowledgements}
This research is supported by Rhode Island Foundation (Grant No.
20144261). The authors would like to appreciate the kind support
of the patients and the healthy individuals who took part in this
study.
%\vspace{-2mm}
%\section*{Acknowledgment}
%The authors would like thank the participants for their co-operation in data collection.
%\nocite{*}
%\begin{thebibliography}
\bibliographystyle{IEEEtran}
\bibliography{echowear_arxiv}
%\bibliographystyle{ieeetran} 
%\bibliography{bigdata}
%\end{thebibliography}
\end{document}